\documentclass[
nofootinbib,
 amsmath,amssymb,
 aps,
]{revtex4-1}
\usepackage{amsmath}
\usepackage{amssymb}
\usepackage{bm}
\usepackage{color,graphicx}
\usepackage{slashed}
\usepackage{comment}
\usepackage{float}
\usepackage{stmaryrd}
\usepackage{cancel}
\usepackage{hyperref}

\begin{document}

\title{Brownian GPDs: A Stochastic Calculus Approach to Generalized Parton Distributions from Integral Constraints}

\author{Douglas Adams}
\affiliation{Department of Physics, University of Virginia, Charlottesville, VA 22904, USA}
\email{yax6jr@virginia.edu}

\author{Simonetta Liuti} 
\affiliation{Department of Physics, University of Virginia, Charlottesville, VA 22904, USA}
\email{sl4y@virginia.edu}

\author{Zaki Panjsheeri}
\affiliation{Department of Physics, University of Virginia, Charlottesville, VA 22904, USA}
\email{zap2nd@virginia.edu}

\author{Brannon Semp}
\affiliation{Department of Physics, University of Virginia, Charlottesville, VA 22904, USA}
\email{hhu8ne@virginia.edu}

\begin{abstract}
We present a stochastic calculus approach for modeling generalized parton distribution functions using information provided from lattice QCD calculations of Mellin moments, without assuming a functional form. Reconstructing the underlying dependence on the partonic momentum fraction from a limited number of moments constitutes an intrinsically non-unique inverse problem. Our goal is to provide a model agnostic way to determine GPDs subject to integral constraints. 
Thus, we derive and show usage of a controlled Langevin approach based on stochastic calculus and Brownian motion.
The method combines a variational formulation, based on minimizing a Dirichlet action subject to the imposed constraints, with stochastic dynamics that generates an ensemble of admissible functions. The resulting ensemble provides a preferred reconstruction and characterizes the variability associated with incomplete integral information within the chosen stochastic model. The framework can accommodate general integral kernels and prescribed function values, offering a flexible approach to reconstructing functions from indirect constraints. Beyond its application to GPD phenomenology, the method provides a general strategy for investigating inverse problems in which integral observables are known but the underlying functional dependence remains undetermined.
The approach may have more general applications beyond the scope of studying 3D hadron structure, and it is particularly useful in situations where we know definite integrals of functions with kernels, but do not know the underlying function. 
\end{abstract}

\maketitle

\section{Introduction}
The extraction of generalized parton distributions (GPDs) from experimental observables constitutes an inverse problem in which the underlying functions are constrained indirectly through integral quantities, the Compton form factors (CFFs). GPDs encode multidimensional information on hadron structure, providing access to the angular momentum carried by quarks and gluons through the angular momentum sum rules \cite{Jaffe:1989jz,Ji:1996ek}, as well as to tomographic images of their spatial distributions inside the proton (see reviews \cite{Diehl:2003ny,Belitsky:2005qn,Liuti:2026jid} and references therein).
An additional source of information is provided by the Mellin moments of GPDs, which are related to generalized form factors (GFFs) accessible through lattice QCD (LQCD) calculations. GPDs satisfy the polynomiality condition, which constrains the dependence of their Mellin moments on the skewness $\xi$, with coefficients given by GFFs that depend on the momentum transfer, $t$. Nevertheless, reconstructing the underlying dependence on the partonic momentum fraction $x$ from a finite number of moments remains an intrinsically nonunique inverse problem.

An important question is whether information contained in one class of integral projections, such as the CFFs, can be used to constrain another, such as the Mellin moments, without requiring a detailed reconstruction of the underlying GPD. This is particularly relevant for determining the total angular momentum carried by quarks and gluons, which is related through the Ji sum rule \cite{Ji:1996ek}, to the second Mellin moment of the GPD combination, $H+E$, in the forward limit. Establishing a connection between different integral projections could therefore provide a more direct way of extracting physically relevant quantities from experimental data, while reducing the dependence on assumptions about the full $x$-dependence of the GPDs. 

As a first step toward this goal, this paper develops a stochastic framework for reconstructing functions from a finite set of integral constraints, using a controlled Langevin process based on Brownian motion. Rather than introducing a specific functional parametrization for the GPD, the method generates an ensemble of stochastic paths conditioned on the available moment information. The resulting ensemble provides a preferred reconstruction and characterizes the variability of the functions compatible with the imposed constraints within the chosen stochastic model.
Brownian motion is a continuous stochastic process characterized by independent Gaussian increments. By introducing a controlled drift into the corresponding stochastic differential equation, the trajectories can be conditioned to satisfy prescribed integral constraints. The stochastic forcing is not introduced as an uncertainty added to a predetermined solution. By contrast, it provides a mechanism for exploring alternative functional reconstructions while maintaining the imposed constraints. Importantly, the stochastic trajectories represent alternative mathematical reconstructions of the underlying distribution, and are not representative of physical Brownian motion of partons inside the proton.

The present study establishes the mathematical formulation and numerical implementation of this approach for moment-constrained reconstruction. In a subsequent extension we will introduce the CFFs, investigating how the information extracted from CFFs can be propagated to probability distributions for Mellin moments, the underlying GPD being treated as a ``latent stochastic function". This would provide a framework for studying the relation between different experimentally and theoretically accessible integral projections without requiring a unique reconstruction of the full GPD.

This paper is organized as follows: in Section \ref{sec:derivation_Langevin} we describe the ingredients of our stochastic reconstruction framework, starting from the constraints given by the Mellin moments of GPDs, and we derive the corresponding Langevin formulation, also describing the essential steps in the computational part ; in Section \ref{sec:results} we present numerical results for GPD extraction from moments, emphasizing the specific kinematic constraints; in Section \ref{sec:conclusions} we draw conclusions and describe future possible extensions and goals.

\section{Langevin Brownian Control}
\label{sec:derivation_Langevin}
The inverse problem we address is to determine an unknown function, given a finite set of integral constraints. 
In the application to GPDs, these constraints are provided by the Mellin moments and Compton form factors that we define below.
The stochastic trajectories generated by the Langevin equation should not be interpreted as the physical motion of partons inside the proton. While quantum fluctuations and correlations are intrinsic to proton structure, identifying the stochastic reconstruction trajectories with such physical fluctuations would require an additional dynamical connection to QCD. 
The evolution variable considered here is the partonic momentum fraction $x$, rather than physical time, and the Brownian increments provide a means of sampling different functions consistent with the imposed integral constraints. 
The resulting ensemble characterizes the {\it variability} of the reconstructed GPDs within the assumed stochastic framework. The latter cannot be immediately identified with the uncertainty, although it allows its quantitative evaluation as we explain in what follows.  

\subsection{Symmetries and QCD Mellin Moments}
\label{subsec:moments}
We begin with describing the integral constraints considered in our approach, which are given by the Mellin moments and the Compton form factors (although the present version of the GPD reconstruction does not make use of the latter). The Mellin moments are defined as,
\begin{equation}
M_n(\xi,t;Q^2)=\int_{-1}^{1} dx\,x^{n-1} H(x,\xi,t;Q^2),
\end{equation}
which can, for example, be calculated in LQCD. We restrict our calculation to the GPD, $H$, written as a function of $x$, the longitudinal parton momentum fraction, $\xi$ the skewness parameter, and the Mandelstam invariant $t$ describing the momentum transfer squared between the intial and final partons in the QCD correlation function; the GPD depends also on the QCD scale, $Q^2$ that we will address later on in this paper. A similar definition can be given for the GPD, $E$. We do not address other GPDs defined in the axial vector, chiral-odd, and gluon sectors, which obey different symmetry relations. The present work can be easily extended to these sectors at a later moment.   
The polynomiality property of GPDs states that the Mellin moments of $H_q$ ($q$ being the quark flavor)are expressible as polynomials in $\xi$, 
\begin{eqnarray}
\label{eq:H_moment}
    \int_{-1}^{1} dx x^{n-1} H_q(x, \xi, t) & = & \sum_{i=0,even}^{n-1} (2\xi)^i A^q_{n,i}(t)
    + \mod(n-1,2) (2\xi)^{n} C_{n}^q(t), 
\end{eqnarray}
and the values of the generalized form factors $A, B, C$ can be computed in LQCD (see Ref.\cite{Cichy:2026xrh} for a review). 

Notoriously, a finite number of moments does not uniquely specify the $x$ dependence of the underlying GPD \cite{Yndurain:1999ui}. 
Reconstructing $H(x,\xi,t)$ therefore requires additional assumptions or a prescription for exploring the set of functions consistent with the imposed constraints.
The reconstruction of GPDs from incomplete integral information has been addressed so far in various approaches, including phenomenological parametrizations \cite{Vanderhaeghen:1999xj,Goloskokov:2007nt,Hannaford-Gunn:2024aix,Panjsheeri:2025vpa}, polynomial and conformal-moment expansions \cite{Kumericki:2011zc,Mamo:2024vjh, Guo:2025muf} and, more recently, neural-network representations \cite{Dutrieux:2021wll,Xu:2026lko} and interpretable machine learning methods like symbolic regression \cite{EXCLAIM:2025iff}. 

\begin{figure}[!h]
  \centering
{\includegraphics[width=0.45\textwidth]{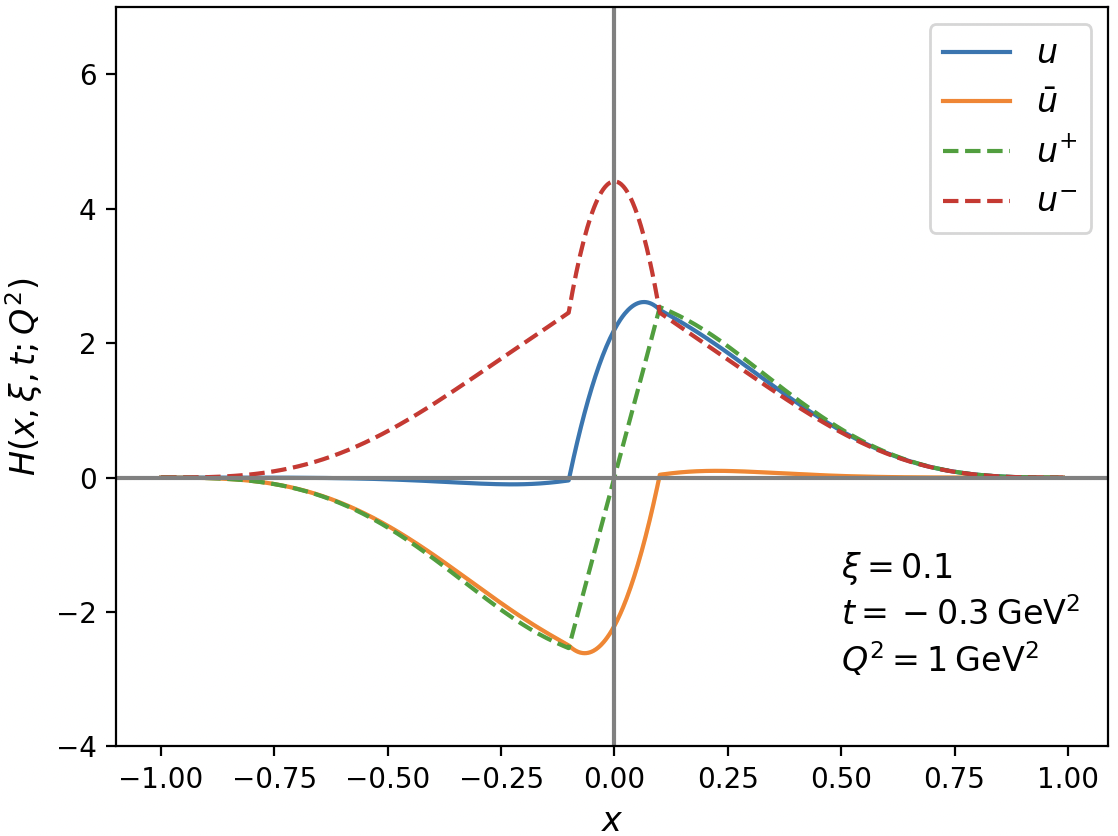}}
{\includegraphics[width=0.45\textwidth]{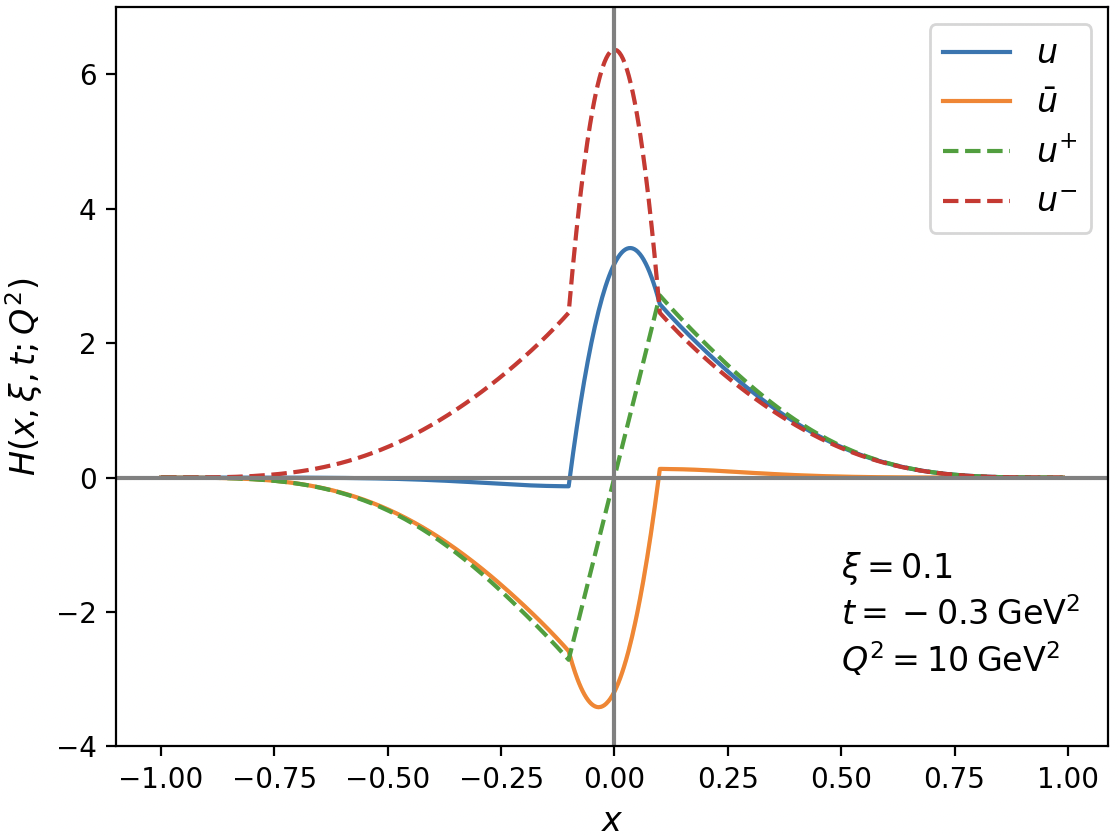}}
  \hfill
  \caption{Typical behavior of the GPD, $H$, illustrating the symmetric  component, $H^-$, the antisymmetric component, $H^+$ and the linear combinations of the two, giving the $u-$quark, $q=u$, and antiquark, $\bar{q}= \bar{u}$ distributions, respectively, at (left) $Q^{2} = 1$ GeV $^{2}$ and (right) $Q^{2} = 10$ GeV$^{2}$ for $\xi = 0.1, t = -0.3$ GeV$^{2}$. The axis of symmetry is at $x = 0$. $H$ was evaluated using the parametrization of Ref.\cite{Panjsheeri:2025vpa}.}
\label{fig:symmetries}
\end{figure}
\begin{figure}[!h]
  \centering
{\includegraphics[width=0.5\textwidth]{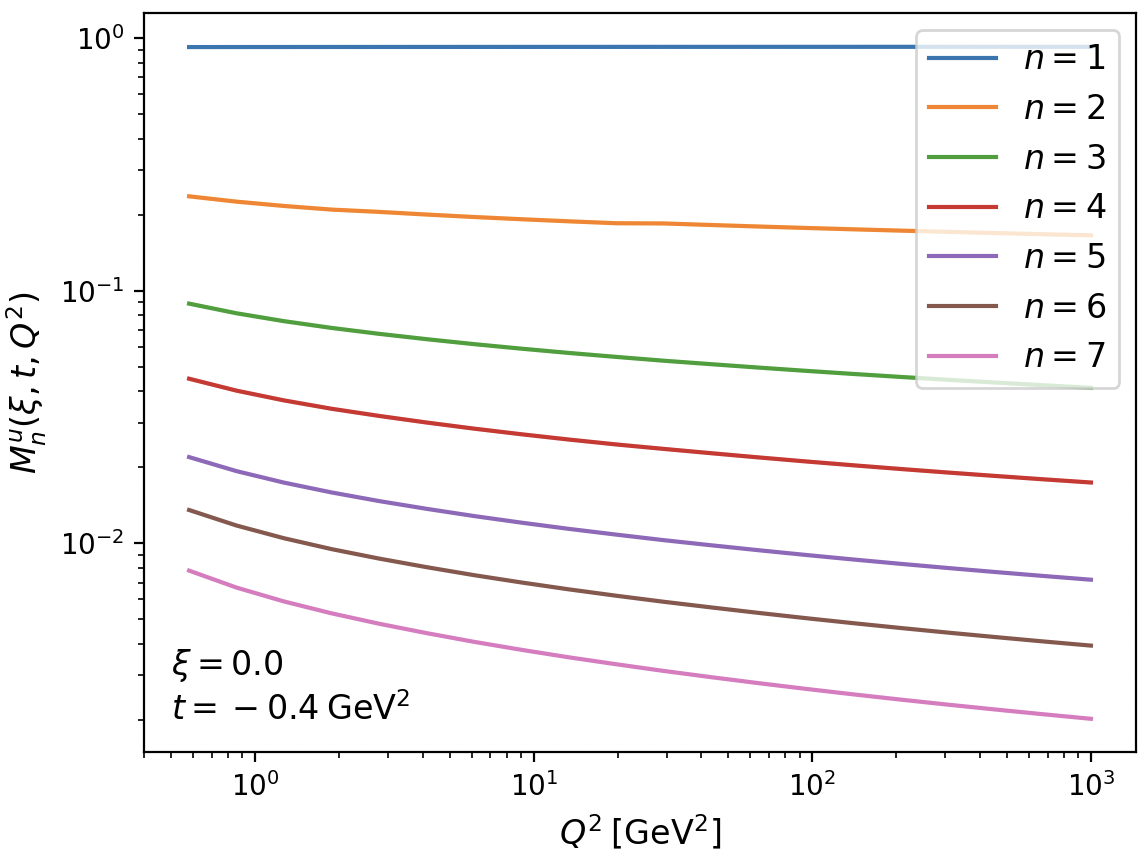}}
  \hfill
  \caption{$Q^2$ dependence of the moments of the GPD, calculated at LO in PQCD (\cite{Panjsheeri:2025vpa, Bertone:2022frx}) for the $u-$quark distribution, using the parameterization of  \cite{Panjsheeri:2025vpa}.  }
\label{fig:Q2dep}
\end{figure}

Here, rather than introducing an explicit functional parametrization or introducing a neural network, we formulate the problem stochastically. 
For fixed values of the external kinematic variables $(\xi,t)$, we render the unknown function through a stochastic path,
\begin{equation}
y(x)\equiv H^{q}(x,\xi,t),
\end{equation}
where the role of the ``evolution" 
\footnote{To be clear, the evolution here refers to the generation of the stochastic path, not perturbative QCD evolution.}
variable is played by $x$. The objective is then to generate an ensemble of paths that satisfy the available integral constraints. Recognizing the quark-antiquark symmetry $H^{q}(x) = -H^{\bar{q}}(-x)$, dictated by charge conjugation, one can show that the {\it even} Mellin moments project out the ``$+$" distribution and the {\it odd} Mellin moments project out the ``$-$" distribution, 
\begin{align*}
M_{n}^{q} (\xi, t) &= 
\left\{
    \begin{array}{lr}
         \int_{0}^{1}dx x^{n-1}H^{+}_{q}(x, \xi, t) \quad \quad \text{(even $n$)}\\
         \\
         \int_{0}^{1}dx x^{n-1}H^{-}_{q}(x, \xi ,t), \quad \quad \text{(odd $n$)}
    \end{array}
\right.,
\end{align*}
where, $n=1,2,3,...$,  $H_{q}^{+} \equiv H^{q} + H^{\bar{q}}$ and $H_{q}^{-} \equiv H^{q} - H^{\bar{q}}$. The symmetries of the distribution are demonstrated in Figure \ref{fig:symmetries} using the parametrization from Ref.\cite{Panjsheeri:2025vpa} for a a specific, typical kinematic point, namely, $\xi=0.25$, $t=- 0.3 $ GeV$^2$, and $Q^2 = 4 $ GeV$^2$. The values of the Mellin moments at the same kinematic point are shown in Figure \ref{fig:Q2dep}, as a function of $Q^2$, as predicted in LO PQCD. 

Therefore only half of the moments from LQCD can be used to reconstruct the GPD with a given symmetry constraint at $x=0$: the $M_1, M_3, M_5 ... $ series for the $-$ distribution and the $M_2, M_4, M_6 ....$ series fopr the $+$ distribution. However, assuming the moments were known independently, for instance integrating a parametric form in $x$, it is mathematically correct to use all of the moments, without distinguishing even and odd, on the   $x \in [0, 1]$ interval. 
Both options were explored in this study with the stochastic model, in which the $+$ distribution is reconstructed over the interval $x \in [0, 1]$, thereby discarding half the moment information, and in which the $q$ distribution is reconstructed over the interval $x \in [-1, 1]$, where all the moments contribute. 

Finally, we observe that the GPDs enter the experimental observables in integral forms, convoluted  with  hard-scattering kernels perturbatively calculable in perturbative QCD, denoted as Compton form factors (CFFs). Even if we do not implement them in the present computation,  we give below an example of 
a CFF for the deeply virtual Compton scattering process (DVCS), for future reference.
At leading twist, and leading order in perturbative QCD (PQCD),  the CFF, ${\mathcal F}_q$, associated with a quark GPD, $F_q = H_q, E_q$, is defined as, 
\begin{equation}
\label{eq:CFF}
\mathcal{F}_q(\xi,t,Q^2)
=
\int_{-1}^{1} dx\,
C^{+}(x,\xi)\,
F_q(x,\xi,t;Q^2),
\end{equation}
where $C^{+}(x,\xi)$ is the Wilson coefficient function at lowest order. Explicitly,
\begin{eqnarray}
 \label{eq:WCF}
 { C^\pm(x,\xi) = \frac{1}{x-\xi + i \epsilon} + \frac{1}{x+\xi - i \epsilon}  } , 
\end{eqnarray}
The real and imaginary parts of the CFFs can be therefore separated out as,
\begin{eqnarray}
    \mathcal{F}_q(\xi,t)  &=&  P.V. \int_{-1}^{1} dx  \left( \frac{1}{x-\xi} + \frac{1}{x+\xi} \right)  F_q(x,\xi,t) 
     {- i \pi \, \big(F_q(\xi,\xi,t) - F_q(-\xi,\xi,t)\big)}\, .
\end{eqnarray}
The CFFs being antisymmetric for $x \leftrightarrow -x$, in this case involve only the GPD $H^+$ which is antisymmetric around $x=0$.

\subsection{Langevin method}
\label{subsec:Langevin}
To incorporate the integral constraints directly into the stochastic evolution, we define an augmented state space, which includes the path, $y$, and also includes new dimensions, $z$, which are dependent upon the path, $y$,
\begin{equation}
X_x =
\begin{bmatrix}
y(x)\\
z_1(x)\\
z_2(x)\\
\vdots\\
z_N(x)
\end{bmatrix},
\label{eq:X}
\end{equation}
where the, $z$, values are running moments as a path is drawn
\begin{equation}
z_i(x)=\int_{x_0}^{x} dx'\,K_i(x')\,y(x') .
\label{eq:running_moments}
\end{equation}
$z_i(x)$ accumulates the $i$th known integral constraint. For Mellin moments, the kernels are simply $K_i(x)=x^{i-1}$. At the endpoint, $x_f$, the components $z_i(x_f)$ are constrained by the known moments, $X_f$.

We then perform a stochastic reconstruction can then be written in a form known as the linear controlled augmented Langevin equation,
\begin{eqnarray}
\label{eq:controlled_langevin}
\boxed{
dX_x =
  \underbrace{
    \bigg( A(x)X_x + B B^T C(x)^{-1} (a_{\text{target}}(x) - X_x) \bigg)
    }_{ f_{\text{drift}    } (X_x, x) } \ \ dx  
+ \underbrace{ \bigg( B \bigg)  
    _{ g_{\text{diffusion}} (X_x, x) } \ \ dW_x \, .}
    }
\end{eqnarray}

The first term, $f_{\text{drift}}$, describes the deterministic evolution associated with the augmented variables, while $dW_x$ generates Brownian fluctuations, $g_{\text{diffusion}}$, in the unknown function. Notice that the additional drift proportional to
\begin{equation}
\label{eqn:control}
C^{-1}(x)\left(a_{\rm target}(x)-X_x\right)
\end{equation}
acts as a control: it biases otherwise stochastic trajectories toward the subset of paths that satisfies the prescribed integral constraints. The matrix $C(x)$ determines how strongly this conditioning acts as the endpoint is approached. 

Eq.~(\ref{eq:controlled_langevin}) therefore provides the central object of our construction. Individual realizations correspond to different functions compatible with the imposed information, while their ensemble defines a probability distribution over possible reconstructions. The conditional mean $a_{\rm target}$ identifies the mean reconstructed path, whereas the spread of the stochastic trajectories quantifies the non-uniqueness, or variability, not fixed by the available finite set of moments.
The resulting posterior distribution provides a basis for reporting credible intervals or uncertainty bands. According to the Bernstein--von Mises theorem, in fact,  the posterior distribution can be approximated by a normal distribution centered at the maximum likelihood estimator thus validating the credible intervals from our analysis as optimal confidence regions. However, to determine a full statistical interpretation of the latter and the uncertainty of the underlying GPD, would require, in principle, a further analysis including the prior assumptions and likelihood (see {\it e.g.} Ref.\cite{Giordano:2020} and references therein). 

Finally, the form of Eq.~(\ref{eq:controlled_langevin}) is not introduced phenomenologically. It follows from a variational principle in which Brownian trajectories are conditioned by known integral constraints, not from physical motion. 
We show this connection in the following section.

\subsection{Variational origin of the controlled Langevin equation}
\label{sec:action}
The stochastic reconstruction described in Section \ref{subsec:Langevin}, derives from a variational formulation in which an action functional assigns a cost to candidate paths subject to the imposed integral constraints, given in this case by distance to a target path, which exactly obeys the desired Mellin moments. 
Minimization of this action identifies a preferred reconstruction, while the associated controlled Langevin dynamics generates an ensemble of paths conditioned on the same constraints. The action therefore plays a practical role in resolving the nonuniqueness of the inverse problem by specifying how different admissible functions are weighted. We remark that this variational principle governs solely the reconstruction procedure rather than any physical dynamics of partons inside the proton.

 For an unconstrained one-dimensional path $y(x)$, the corresponding Dirichlet action is
\begin{equation}
S_0[y]
=
\frac{1}{2}\int_{x_0}^{x_f}dx
\left(\frac{dy}{dx}\right)^2 
\end{equation}
This action penalizes rapid variations of the path and provides the quadratic functional underlying the Brownian process. The integral information can then be incorporated through the augmented state $X_x$ introduced above and by conditioning the endpoint on the prescribed values of the moments.

For the augmented system, we introduce the natural drift $A(x)$ and controllability vector $B$. The constrained action can then be written as

\begin{equation}
\label{eqn:action_aug}
S_{\rm aug}
=
\int_{x_0}^{x_f}dx\,
\left[
\frac{1}{2}
\left\|
B^\dagger
\left(
\dot X_x-A(x)X_x
\right)
\right\|^2
+
\frac{1}{2\sigma^2}
\left(X_x-a_{\rm target}\right)^T
\left(X_x-a_{\rm target}\right)
\right],
\end{equation}

where $\sigma$ controls the strength with which the constraints are imposed. In the limit of a sharply specified endpoint, $\sigma\rightarrow0$, trajectories are conditioned to satisfy the prescribed integral constraints.

For the linear-Gaussian problem considered here one can show using the Hamilton--Jacobi--Bellman equation determines $C(x)$ through the associated Riccati equation. Minimization yields a control:


 \begin{equation}
u^\ast(x)
=
BB^T C^{-1}(x)
\left(a_{\rm target}(x)-X_x\right),
\end{equation}
which, when inserted into the stochastic dynamics,
\begin{equation}
dX_x=\left[A(x)X_x+ u^\ast(x)\right]dx+B\,dW_x,
\end{equation}
reproduces the controlled Langevin equation, Eq.(\ref{eq:controlled_langevin}).

\subsection{Computation}
\label{subsec:Compute}
The stochastic differential equation provides a numerical realization of the variational reconstruction framework introduced in the preceding section. The action functional (Sec.\ref{sec:action}), specifies a preferred path subject to the imposed moment constraints, while the controlled stochastic dynamics generates an ensemble of admissible reconstructions. 
The evolution parameter is the partonic momentum fraction $x$. 
The numerical implementation of the stochastic reconstruction framework combines the natural evolution of an augmented dynamical system with a controlled Langevin process that enforces the available integral constraints, Eq.\eqref{eq:controlled_langevin}. 
We first introduce the matrix forms for the drift, ($A$), and diffusion, ($B$ and $G$) operators, the propagator as a function of $A$, and the controllability Gramian, $W$, which determine how the trajectories evolve and are conditioned by the Mellin moments. We then describe the Euler–Maruyama discretization used to generate an ensemble of stochastic paths consistent with these constraints.

\subsubsection{Using Langevin equation to solve for the best path}
The Natural Drift ($A$) describes how a path would evolve over time if no control and no randomness is applied to a path (unforced system evolution).  We can see this intuitively by writing the differential equation in stochastic form with, $dX(x) = A \, X(x) dx$, where we ignored the control and the randomness of the Brownian motion, and therefore the natural drift matrix is given by $A$. 
The matrix $A$ is defined as,
\begin{eqnarray}
\label{eq:Amatrix}
A = 
\begin{bmatrix} 
0 && 0 && \cdots && 0 \\
K_1(x) && 0 && \cdots && 0 \\
\vdots && \vdots && \ddots && \vdots \\
K_j(x) && 0 && \cdots && 0 \\
\end{bmatrix}
\end{eqnarray}
$A$ keeps track of the running integrals with kernels $K_i$, in a system of linear equations.
The top row of the natural drift being zero implies that without any forcing, the plain path $y$ which is the first element of $X$ does not evolve. 
The integrals with the kernels give the drift, represented by $z_i$, as defined in Eq.\eqref{eq:running_moments}.

\vspace{0.5cm}
\noindent \underline{Controllability Vector ($B$) and the associated ``Diffusion Tensor/Noise Intensity Matrix" ($G$)}

\noindent The vector $B$ and matrix, $G$, in Eq.\eqref{eq:controlled_langevin}, specify that the stochastic forcing acts directly only on the primary path, $y(x)$, while all the other paths, through running moments ($z_i(x)$), Eq.\eqref{eq:running_moments}, are determined by the running integrals of $y(x)$, and depend on the kernels, $K_i(x)$.
The control term, $BB^T C^{-1}(x)(a_{target} - X_x)$, only acts upon the first element of $X$, Eq.\eqref{eq:X} which is the path $y$. 
The vector, $B$, 
\begin{eqnarray}
B = [1, 0, \cdots, 0]
\end{eqnarray}
and matrix, $G$, 
\begin{eqnarray}G = B B^{\top}= \begin{bmatrix} 
1 && 0 && \cdots && 0 \\
0 && 0 && \cdots && 0 \\
\vdots && \vdots && \ddots && \vdots \\
0 && 0 && \cdots && 0 \\
\end{bmatrix}
\end{eqnarray}
serve the purpose of isolating control over the path $y$ in our system of equations, while admitting there is no independent control over any of  $z_i$. 

\vspace{0.3cm}\noindent \underline{Linear time variant (LTV) propagator (Matrix Exponential Form)}

\noindent The propagator is the mathematical operator that takes a state at ``time", $x$ and evolves it forward to time $s$, under the system's ``natural" dynamics, 
-- the solution to the unforced system $d X = A X dx$. It is written in regular form as 
\cite{Hespanha2018-pk} 
\cite{Chen1998-iz},
\begin{eqnarray}
\Phi(s, x) = \exp\left( \int_{x}^{s} A(\tau) \, d\tau \right) .
\end{eqnarray}
To approximate the exponential of a matrix, we adopt the Peano-Baker series 
which consists in the following expansion,
\begin{eqnarray}\Phi(s,x ) = I + \int_{x}^{s} A(\tau_1) d\tau_1 + \int_{x}^{s} A(\tau_1) \int_{x}^{\tau_1} A(\tau_2) d\tau_2 d\tau_1 + \int_{x}^{s} A(\tau_1) \int_{x}^{\tau_1} A(\tau_2) \int_{x}^{\tau_2} A(\tau_3) d\tau_3 d\tau_2 d\tau_1 + \dots
\end{eqnarray}
We take a ``nilpotent shortcut" noticing that the matrix $A$ contains mostly 0's, which is a standard linear expansion trick, to reduce the number of terms kept in the Peano-Baker series. In our case, Eq.\eqref{eq:Amatrix}, the higher order terms in the series are null since all the elements of the $A$ matrix are zero outside of the first column, yielding,
\begin{eqnarray}
\Phi(s, x) = I + \int_{x}^{s} A(\tau) \, d\tau
\end{eqnarray}

\vspace{0.3cm}
\noindent \underline{Gramians: $W(x_a, x_b)$, $C_{fixed}(x)$, $C_{prob}(x)$}

\noindent The general controllability Gramian, $W(x_a, x_b)$, finds the covariance matrix of the state $X$ caused by accumulated continuous noise, $Cov(X) = E[X X^T]$ -- $X X^{T}$ denotes a Gramian \cite{Chen1998-iz}. 
The stochastic trajectories are driven toward a prescribed final state -- the target trajectory --  while Brownian noise continuously perturbs their evolution along its walk.
The effect of all these fluctuations must therefore be accounted for when determining the control required to satisfy the endpoint constraints, while assessing how much they are working against the control. 
This is achieved through the controllability Gramian, which integrates the contributions of the stochastic forcing along the trajectory, propagated to the final state by the natural evolution operator, $\Phi$, 
\begin{eqnarray}
W(x_a, x_b) = \int_{x_a}^{x_b} \Phi(x_b, \tau) \, G \, \Phi(x_b, \tau)^\top d\tau
\end{eqnarray}
The stochastic fluctuations introduced at different values of $x$ contribute to the covariance of the final state,

\noindent The fixed-endpoint bridge covariance Gramian, $C_{fixed}(x)$, is the same Gramian, implementing the final path location $x_f$   \cite{Stengel2012-cf} 
\begin{eqnarray}
C_{fixed}(x) = W(x, x_f) = \int_{x}^{x_f} \Phi(x_f, \tau) \, G \, \Phi(x_f, \tau)^\top d\tau
\end{eqnarray}
$C_{fixed}(x)$ gives the covariance of the stochastic fluctuations accumulated between the current position $x$, and the prescribed final position $x_f$, propagated to the endpoint by the natural dynamics.

\noindent 
The probabilistic-endpoint Gramian, $C_{\rm prob}(x)$, extends the fixed-endpoint formulation by incorporating uncertainty in the prescribed terminal constraints, characterized by the covariance matrix $\Sigma_{X_f}$ \cite{Oksendal1995-wg},
\begin{eqnarray}
C_{prob}(x) = W(x, x_f) + \Phi(x, x_f) \, \Sigma_{X_f} \,  \Phi(x, x_f)^\top  .
\end{eqnarray}

\vspace{0.3cm}
\noindent \underline{Fixed-Endpoint Conditional Mean Path, $a_{\text{target}}(x)$ \cite{Anderson1979-mv}}

\noindent This is simply the ``best" path that minimizes the action as defined previously in Eq.\eqref{eqn:action_aug}. The $a_{target}$ path has the least amount of accumulated derivative changes, while also fully satisfying all the moments. It is used as a ``target" for the control terms, to make use of when steering a walker towards desired final moments. We can see this because a difference is shown in the "control" term, Eq. \eqref{eqn:control}.
\begin{eqnarray}
a_{\text{target}}(x) = W(0, x) \Phi(x, x_f)^\top W(0, x_f)^{-1} X_f . 
\end{eqnarray}

Terminal-Augmented Conditional Mean Target Path, is the same target path as above, adjusted for randomness included in the desired moments. 
$\Big(a_{\text{target}}(x)\Big)$ 
\cite{2016ITAC...61.1158C}
\begin{eqnarray}
a_{\text{target}}(x) = W(0, x) \Phi(x, x_f)^\top \left[ W(0, x_f) + \Sigma_{X_f} \right]^{-1} X_f
\end{eqnarray}

\subsubsection{Using Langevin equation to generate Brownian paths}
\label{subsec:ito}
To generate stochastic trajectories numerically, we use the Euler–Maruyama method, which discretizes the Langevin equation into a sequence of small steps \cite{KloedenPlaten1992,Oksendal2003}. At each step, the trajectory is updated by combining a deterministic contribution from the drift with a random contribution from Brownian motion. This procedure follows the It\^{o} formulation of stochastic differential equations, in which the drift and diffusion coefficients are evaluated at the beginning of each step. Notice that in the present framework, the diffusion coefficient, $g_{\rm diffusion}=B$, is, however,  constant and does not depend on the state $X(x)$. 

A defining property of Brownian motion is that increments over non-overlapping intervals are independent Gaussian random variables, with
\begin{equation}
\langle\Delta W_i\rangle=0,
\qquad
\langle(\Delta W_i)^2\rangle=\Delta x .
\end{equation}
Consequently,
\begin{equation}
\Delta W_i\sim{\cal N}(0,\Delta x),
\end{equation}
and in the numerical implementation it can be generated according to
\begin{equation}
\Delta W_i=\sqrt{\Delta x} \, Z_i,
\qquad
Z_i\sim{\cal N}(0,1).
\end{equation}
An important feature of stochastic differential equations is the different scaling of the deterministic and stochastic contributions. Over a small interval $\Delta x$, the deterministic drift produces a displacement proportional to $\Delta x$,
\begin{equation}
\Delta X_{\rm drift}
\sim
f_{\rm drift}\Delta x,
\end{equation}
whereas the typical stochastic displacement scales as
\begin{equation}
 \Delta X_{\rm stochastic}\sim g_{\rm diffusion}\sqrt{\Delta x}.
\end{equation}
Although both contributions vanish as, $\Delta x \to 0$, Brownian trajectories are not differentiable in the ordinary sense. This behavior motivates the use of stochastic calculus and the Euler-Maruyama discretization.

Applying this procedure to the controlled Langevin equation gives,
\begin{equation}
X_i
+
\left[
A(x_i)X_i
+
BB^T C^{-1}(x_i)
\left(
a_{\rm target}(x_i)-X_i
\right)
\right]\Delta x
+
B\sqrt{\Delta x} \, Z_i .
\label{eq}
\end{equation}
Eq.~(\ref{eq}) makes the numerical construction particularly transparent. At every step, the trajectory first receives a deterministic displacement determined by the drift. Part of this drift drives the natural evolution of the augmented variables, while the control term biases the trajectory toward satisfying the imposed integral constraints. Note however, under the current construction, the moments may not be obeyed exactly, because the control is guided towards the optimal target path, instead of a full projection forward at each step. A random Gaussian displacement is then added through the Brownian term.

Different sequences of the random variables
\begin{equation}
{Z_0,Z_1,\ldots,Z_{N-1}}
\end{equation}
generate different stochastic trajectories. Repeating the procedure therefore produces an ensemble of possible paths rather than a single reconstructed function. The paths fluctuate because of the Brownian term, while the controlled drift ensures that these fluctuations remain conditioned to be close to integral information by being near the target path.

The role of the control becomes increasingly important as the constrained endpoint is approached. The covariance $C(x)$ characterizes the remaining freedom of the stochastic trajectory, and its inverse $C^{-1}(x)$ increases the restoring influence of the target as this freedom decreases. The resulting process can therefore explore different trajectories at intermediate points while being progressively driven toward the prescribed constraints.

The Euler–Maruyama method thus provides a numerical procedure for generating an ensemble of reconstructed paths from the continuous Langevin equation. 
At every step we add the deterministic drift, and a random diffusion to the walker. 
The deterministic drift, includes the control needed to help the walker get close to hitting the final target. 
The control gets more powerful as the walker gets further along the path, whereby at each step, the state is updated according to
\begin{eqnarray}
X_{i+1} = X_{i} + \text{f}_{drift} (\Delta x) + \text{f}_{ diffusion} (\Delta W)
\end{eqnarray}
where $\Delta W_i=\sqrt{\Delta x} Z_i$, with $Z_i\sim\mathcal N(0,1)$. The deterministic drift includes both the natural evolution and the control required to impose the endpoint constraints, while the Brownian term generates different stochastic realizations.

The resulting ensemble provides a representation of the GPD functions compatible with the imposed moment constraints, allowing the variability of the reconstruction to be examined without selecting a single functional parametrization.
\section{Results}
\label{sec:results}
\begin{figure}[!h]
  \centering
{\includegraphics[width=0.55\textwidth]{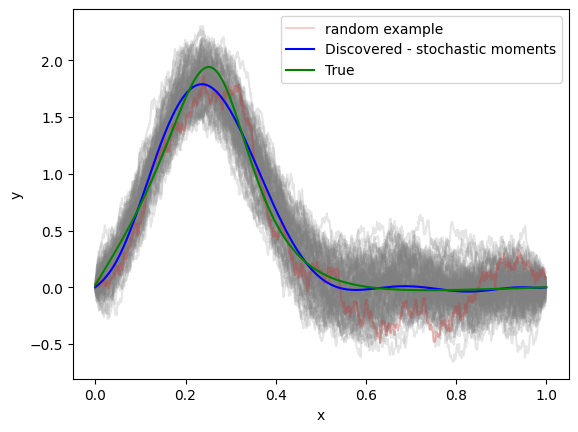}}\label{fig:stoch_lang7m}
  \hfill
  \caption{ 
  \label{fig:example1_Langevin7m}
  Hidden True path and Stochastic  Reconstruction using the first 7 moments as information while assuming the Linear Augmented Controlled Langevin Equation. Each of 100 realized paths is shown in grey. The blue curve is ``best" as it satisfies the moments while also fully minimizing the smoothness term in the action ( Eq. \protect{\ref{eqn:action_aug}} ). The red curve is a randomly chosen instance of the grey curves, to provide a visualization of what one might look like. }
\end{figure}

\begin{figure}[!h]
  \centering
{\includegraphics[width=0.6\textwidth]{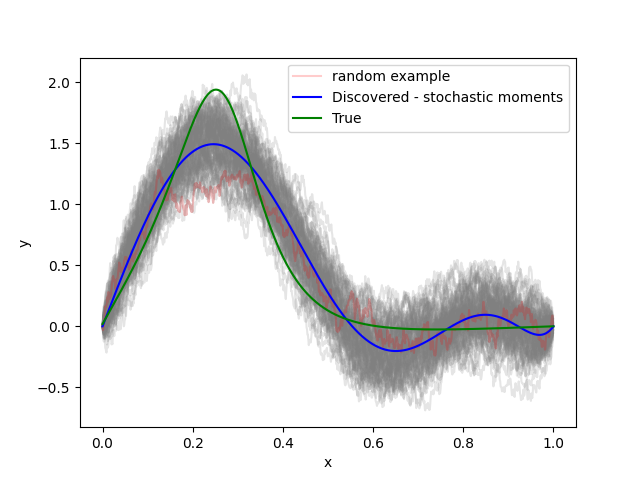}}\label{fig:stoch_lang}
  \hfill
  \caption{ 
  \label{fig:example1_Langevin}
  Hidden True path and Stochastic Reconstruction using the $n-1$ odd-power ones, with $n=2,4,6,8$, Eq.\eqref{eq:H_moment}. Notations as in Fig.\ref{fig:example1_Langevin7m}.}
\end{figure}

Our main results are shown in Figures \ref{fig:example1_Langevin7m} and \ref{fig:example1_Langevin}.
Figure~\ref{fig:example1_Langevin7m} illustrates the stochastic reconstruction of a reference function, the GPD, 
\[  H^+(x,\xi, t , Q^2) , \]
using its first seven Mellin moments as input constraints. 
We underline that we perform a {\it local} analysis, {\it i.e.} point-by-point at fixed for $\xi,t \neq 0$ and $Q^2$. The moments were obtained from the parametrization in Ref.\cite{Panjsheeri:2025vpa}, at a fixed $Q^2= 4$ GeV$^2$ for $\xi = 0.18, t = -0.5$ GeV$^2$. PQCD evolution allows us to change the value of $Q^2$ and study the correlation in the $x$ behavior of the distribution accordingly.  
The reference function is treated as unknown during the reconstruction and is used only to assess the results. The linear augmented controlled Langevin equation generates an ensemble of 100 stochastic trajectories, each close to satisfying the prescribed moment constraints.
The blue curve represents the true minimum-action solution, obtained by minimizing the Dirichlet action subject to the same constraints. It provides a preferred reconstruction according to the smoothness criterion introduced in the variational formulation. The gray band constitutes an ensemble of curves, each one representing individual stochastic realizations, while the red curve highlights one such realization to illustrate how a function can differ from the minimum-action solution.

The spread of the stochastic trajectories demonstrates the non-uniqueness of the reconstruction from a finite number of Mellin moments. Although all trajectories are near the least action solution, their pointwise dependence on $x$ can differ substantially. This behavior reiterates that the agreement at the level of integrated quantities does not necessarily imply a unique determination of the underlying function. The stochastic ensemble makes this ambiguity explicit by providing multiple reconstructions compatible with the same input information.

An important feature of the method is that the stochastic trajectories are generated directly through the conditioned Langevin dynamics, rather than by introducing fluctuations around the minimum-action solution after the reconstruction. The resulting variability therefore reflects the combined effect of the imposed moment constraints and the assumed stochastic dynamics. In particular, the ensemble can be used to investigate the variability of the reconstructed function across different regions of $x$ and the sensitivity of the results to the imposed constraints and stochastic assumptions. The  magnitude of the band should not be interpreted as a $1 \sigma$ experimental uncertainty without an additional statistical treatment of uncertainties in the input moments.
However, the controlled Langevin framework provides an ensemble of admissible functions, which is amenable to a subsequent uncertainty quantification (UQ) analysis.  A quantitative UQ analysis, including the propagation of uncertainties and correlations in the input moments, is beyond the scope of the present study and will be addressed in future work. 

In Figure \ref{fig:example1_Langevin} we use only moments of the $H^+$ distribution, or the $n-1$ odd-power ones, with $n=2,4,6,8$, Eq.\eqref{eq:H_moment}. 
In this case, since the reconstruction is constrained by only a subset of the available Mellin moments, one has additional freedom in the functional form of the distribution. The Langevin framework explores the family of functions consistent with these constraints, while the smoothness term in the action provides additional regularization. This example illustrates how the method performs when the available information is incomplete and how the resulting reconstruction depends on the interplay between moment constraints and the assumed stochastic dynamics.

\section{Conclusions and Outlook}
\label{sec:conclusions}
In this work, we have developed a stochastic framework for reconstructing unknown functions from a finite set of integral constraints, motivated by the inverse problem of determining generalized parton distributions (GPDs). The method combines a variational formulation, in which a preferred reconstructed function minimizes the Dirichlet action subject to the imposed constraints, with a controlled Langevin equation that generates an ensemble of stochastic trajectories. The control incorporates the available integral information into the evolution, while the Brownian component allows different functional realizations to be explored. In this way, the reconstruction is not restricted to a single minimum-Dirichlet-action solution. 

Most importantly, 
 the unknown function is not represented by a neural network but by interpretable equations. The reconstruction is formulated directly in function space through a stochastic differential equation conditioned on the available integral information. The ensemble is generated by independent realizations of the Brownian increments, with the control term restricting the trajectories according to the prescribed constraints. Its variability therefore reflects both the incomplete information supplied by the moments and the assumptions defining the stochastic process. 
Importantly, it defines a probabilistic value as a penalty which limits the most oscillatory shadow GPDs \cite{Bertone:2021yyz}.


The central physics motivation is the inverse problem encountered in studies of deeply virtual exclusive scattering experiments, where both the experimental observables  and lattice QCD calculations provide integral information about the underlying quark and gluon distributions, the GPDs, without uniquely determining their functional dependence. The Mellin moments of GPDs are given by matrix elements of local QCD operators. The experimental observables are cast in terms of integrals over the $x$-dependent GPDs with known kernels in a perturbative QCD factorized scenario. It is well known that a finite number of Mellin moments cannot uniquely determine a GPD as a function of the partonic momentum fraction $x$, at fixed skewness, $\xi$, and momentum transfer, $t$. Different functions can reproduce the same moments while exhibiting substantially different behavior in regions not adequately constrained by the available information. This reconstruction problem has been addressed using flexible neural-network representations of GPDs, including the NNGPD framework, which incorporates available experimental information and theoretical constraints to reduce dependence on predetermined functional forms \cite{Xu:2026lko}, in which probability distributions over network parameters induce probability distributions over the reconstructed functions. In this case, Bayesian neural networks provide an important method to address the uncertainty quantification.  The stochastic approach developed here provides a complementary strategy, in which an ensemble of candidate functions is generated directly from the integral constraints through conditioned Langevin dynamics.

%


In the NNPDF methodology \cite{NNPDF:2014otw,NNPDF:2024dpb,NNPDF:2021njg,Carrazza:2019mzf}, uncertainties are propagated by generating Monte Carlo replicas of the experimental measurements and performing independent fits to these replicas. The resulting ensemble of fitted PDFs represents the uncertainty associated with the input data, conditional on the fitting procedure and its assumptions.  Unlike the Monte Carlo replica method, the present construction does not, by itself, propagate experimental uncertainties into the reconstructed function. Such an interpretation requires an explicit statistical model for the uncertainties and correlations of the input constraints. In this context, the Bernstein–von Mises theorem provides an important theoretical connection between Bayesian posterior distributions and frequentist uncertainty quantification. Extensions of this theorem to linear inverse problems \cite{Giordano:2020} establish conditions under which posterior distributions of suitable functionals become asymptotically Gaussian, with uncertainty estimates that admit a frequentist interpretation. These results highlight the importance of distinguishing the variability generated by a chosen stochastic reconstruction model from statistically calibrated uncertainty arising from the available data. Establishing such a connection for the present Langevin framework, including the effects of finite and correlated moment constraints, provides an interesting direction for future investigations.

In our approach the inverse problem is not limited to identifying a function that reproduces the known moments, but also involves determining how much information about the various domains of its $x$-dependence in its support  interval, can be inferred from those moments. The stochastic ensemble provides a framework for studying the sensitivity of the reconstruction to incomplete moment information and to the assumptions introduced in selecting admissible functions. 

%
%
This framework can be further applied to realistic GPD extractions including lattice QCD moments and Compton form factors with  uncertainties and correlations in the input moments, as well as extending the reconstruction to multiple kinematic variables. 
Alternative numerical implementations, including spectral methods, will be considered in subsequent work.

\begin{acknowledgments}
    This work was completed under the Department of Energy grant DE-SC0024644 and DOE grant DE-SC0016286. D.A., S.L. and Z.P.are also grateful to the organizers of the EIC Theory Group Institute in the summer of 2026 for their hospitality and for many fruitful discussions. We thank all members of the ExclAIm Collaboration for continuously providing a stimulating environment. 
\end{acknowledgments}

\bibliography{references}
\end{document}